\documentclass[11pt,a4paper,aps,preprint,superscriptaddress,nofootinbib]{revtex4-1}
\usepackage{graphicx}
\usepackage{amssymb}
\usepackage{amsmath}
\usepackage{color}
\usepackage[colorlinks=true, pdfstartview=FitV, linkcolor=blue, citecolor=blue, urlcolor=blue]{hyperref}
\usepackage[utf8]{inputenc}

\begin{document}

\title{$R_2$ as a single leptoquark solution to $R_{D^{(*)}}$ and $R_{K^{(*)}}$}
\author{Oleg Popov}
\email{opopo001@ucr.edu}
\affiliation{Physics and Astronomy Department, University of California, Riverside, California 92521, USA\\
Institute of Convergence Fundamental Studies,  Seoul National University of Science and Technology, Seoul 139-743, Korea}

\author{Michael A.~Schmidt}
\email{m.schmidt@unsw.edu.au}
\affiliation{School of Physics, The University of New South Wales, Sydney, NSW 2052, Australia}

\author{Graham White}
\email{gwhite@triumf.ca}
\affiliation{TRIUMF Theory Group, 4004 Wesbrook Mall, Vancouver, B.C. V6T2A3, Canada}

\begin{abstract}We show that, up to plausible uncertainties in BR$(B_c\to \tau \nu)$, the $R_2$ leptoquark can simultaneously explain the observation of anomalies in $R_{K^{(*)}}$ and  $R_{D^{(*)}}$ without requiring large couplings. The former is achieved via a small coupling to first generation leptons which boosts the decay rate $\Gamma(\bar B\to \bar K^{(*)}e^+e^-)$. Finally we motivate a neutrino mass model that includes the $S_3$ leptoquark which can alleviate a mild tension with the most conservative limits on  BR$(B_c\to \tau \nu)$.
\end{abstract}


\maketitle

\section{Introduction}

There have recently been multiple independent anomalous measurements of semileptonic $B$ decays that depart from standard model (SM) predictions. Rare decays into $D ^{(*)}$ mesons show a discrepancy from SM predictions in BABAR~\cite{Lees:2012xj,Lees:2013uzd}, Belle~\cite{Huschle:2015rga,Sato:2016svk,Hirose:2016wfn}, and LHCb~\cite{Aaij:2015yra,Aaij:2017deq}  measurements of the lepton flavor universality (LFU) ratios. New results from Belle combined with measurements from BABAR and LHCb give~\cite{BelleRD*new}\footnote{The best-fit value and error bars have been extracted from the figure on slide 9.}
\begin{eqnarray}
	{\cal R}_{D^{}}= \frac{ \Gamma (\bar{B} \to D^{}\tau \bar{\nu })}{\Gamma( \bar{B} \to D^{}e/\mu \bar{\nu })} = \left\{ \begin{array}{cc} 
   0.299 \pm 0.003 & \text{SM~\cite{Lattice:2015rga}} \\ 
   0.335\pm 0.031 & \text{observed~\cite{BelleRD*new}}
   \end{array} \right. \label{eq:rd}
\end{eqnarray}
and
\begin{eqnarray}
	{\cal R}_{D^{*}}= \frac{ \Gamma (\bar{B} \to D^{*}\tau \bar{\nu })}{\Gamma( \bar{B} \to D^{*}e/\mu \bar{\nu })} = \left\{ \begin{array}{cc} 
0.258 \pm 0.005 & \text{SM~\cite{Tanaka:2012nw}} \\ 
0.298\pm 0.015 & \text{observed~\cite{BelleRD*new}}
\end{array} \right. \label{eq:rdstar}\;.
\end{eqnarray}
When the correlation between the two observables is taken into account the significance of the anomaly is at the $3.1\sigma$ level~\cite{BelleRD*new}. The SM calculation is reliable as it is largely insensitive to hadronic uncertainties which cancel out in the ratios $R_{D^{(*)}}$.

LHCb has similarly found an intriguing deviation from LFU in the semileptonic $B$ meson decays to $K^{(*)}$ mesons. The LFU ratios
\begin{equation}
    R_{K^{(*)}} = \frac{\Gamma(\bar B\to \bar K^{(*)} \mu^+\mu^-)}{\Gamma(\bar B\to \bar K^{(*)}e^+e^-)} \label{eq:RKeqn}
\end{equation}
provide a clean probe of new physics effects because hadronic uncertainties cancel out in the ratios as long as new physics effects are small~\cite{Hiller:2003js,Capdevila:2016ivx,Capdevila:2017bsm}.
LHCb measured the ratios for the dilepton invariant
mass range $1.1\,\mathrm{GeV}^2 < q^2 < 6 \,\mathrm{GeV}^2$. A combination of run I and run II from LHCb gives
\begin{equation}\label{eq:RK}
	R_{K}
	=\left\{ \begin{array}{cc} 1.0003\pm 0.0001 & {\rm SM} \text{\cite{Bobeth:2007dw}}\\ 0.846 ^{+0.06} _{-0.054} {\rm (stat)} ^{+0.016} _{-0.014} {\rm (sys)} &   {\rm observed} \text{\cite{Aaij:2019wad}} \end{array}  \right.
\end{equation}
and
\begin{equation}\label{eq:RKstar}
	R_{K^*} 
	=\left\{ \begin{array}{cc} 1.00 \pm 0.01 & {\rm SM} \text{\cite{Bordone:2016gaq}} \\ 0.716 ^{+0.070} _{-0.057} &   {\rm observed} \text{\cite{BelleRK*new,Aaij:2017vbb}} \end{array}  \right.
\end{equation}
where we combined the LHCb measurement~\cite{Aaij:2017vbb} of $R_{K^*}$ with the new Belle measurement~\cite{BelleRK*new} using the methods described in Ref.~\cite{Barlow:2003sg}. 
Experimental sensitivity to both of these anomalies is expected to improve by orders of magnitude over the next few years and make a potential confirmation of a departure from the SM imminent. 
The measurements are not just quantitatively different from the SM but qualitatively so as well, because the SM has no notable violation of lepton flavor universality. 

The most common explanation for these anomalies is to extend the
SM by leptoquarks (see Refs.~\cite{Buras:2014fpa,Gripaios:2014tna,Pas:2015hca,Barbieri:2016las,Duraisamy:2016gsd,Sumensari:2017ovu,Aloni:2017ixa,Sumensari:2017mud,Hiller:2017bzc,DAmico:2017mtc,Cline:2017aed,Guo:2017gxp,Crivellin:2017dsk,Alok:2017sui,Angelescu:2018tyl,deMedeirosVarzielas:2019okf,deMedeirosVarzielas:2019lgb,Sheng:2018vvm,Balaji:2018zna}
for a leptoquark solution to the $R_{K^{(*)}}$ anomalies, Refs.~\cite{Freytsis:2015qca,Biswas:2018jun,Angelescu:2018tyl,Zhang:2019hth,Aydemir:2019ynb,Mandal:2018kau,Bansal:2018nwp,Iguro:2018vqb} for the $R_{D^{(*)}}$ anomalies and Refs.~\cite{Bauer:2015knc,Fajfer:2015ycq,Bhattacharya:2016mcc,Sahoo:2016pet,Becirevic:2016yqi,Becirevic:2016oho,Li:2016vvp,Chauhan:2017uil,Calibbi:2017qbu,DiLuzio:2017vat,Buttazzo:2017ixm,Cai:2017wry,Crivellin:2017zlb,Muller:2018nwq,Angelescu:2018tyl,Cornella:2019hct,DaRold:2018moy,Schmaltz:2018nls,Fornal:2018dqn,Assad:2017iib,Blanke:2018sro,Becirevic:2018afm,Azatov:2018knx,Azatov:2018kzb,Huang:2018nnq}
for simultaneous explanations). Vector leptoquarks have issues with ultraviolet
(UV) completion and their tendency is to be heavy in UV complete models. Therefore
it is attractive to consider scalar leptoquark solutions to these anomalies.
To date the only known candidate that simultaneously explains both sets of anomalies is the $S_1$ leptoquark~\cite{Bauer:2015knc,Popov:2016fzr,Cai:2017wry}, but it only satisfies $R_{K^{(*)}}$ at $2-\sigma$~\cite{Cai:2017wry}. In this work we show that the $R_2$ leptoquark
can provide a simultaneous solution at $1-\sigma$ consistent with all known constraints.

In addition to the anomalous measurements of $R_{D^{(*)}}$ and $R_{K^{(*)}}$, two other anomalies have
generated interest:
On the one hand, the value of the angular observable $P_5^\prime$~\cite{Descotes-Genon:2013wba,Guadagnoli:2017jcl} and more generally the data of $b\to s\mu\bar\mu$ point to a deviation from the SM~\cite{Descotes-Genon:2015uva}. While these anomalies are
intriguing they are currently less clean signals of new physics due to
large hadronic uncertainties and the difficulty in estimating a signal for the
$P_5^\prime$ anomalies~\cite{Guadagnoli:2017jcl}.  
On the other hand, similar to the LFU ratios $R_{D^{(*)}}$ the LFU ratio $R_{J/\psi}=\Gamma(B_c^+\to J/\psi\tau\nu)/\Gamma(B_c^+\to J/\psi \mu \nu)$ points to a larger branching fraction to $\tau$ leptons compared to muons, but it is still consistent with the SM at $2-\sigma$ due to the large error bars~\cite{Aaij:2017tyk}. 
We therefore leave the consideration of such anomalies to future work. 

The $R_2$ leptoquark has quantum numbers $(3,2,7/6)$ with respect to the SM gauge group 
$\mathrm{SU}(3)\times \mathrm{SU}(2)\times \mathrm{U}(1)$
and has been proposed as a cause of the $R_{D^{(*)}}$ anomalies~\cite{Tanaka:2012nw,Dorsner:2013tla,Sakaki:2013bfa} with O(1) couplings as well
as the $R_{K^{(*)}}$ anomalies with very large couplings through a new contribution to the decay $b\to s \mu\bar\mu$~\cite{Sahoo:2015wya,Chen:2016dip,Dey:2017ede,Becirevic:2017jtw,Chauhan:2017ndd}. These operators are induced at the 1-loop
level and thus require undesirably large couplings with at least one coupling needing
to be a lot larger than $1$. 
We reopen the case of this leptoquark and find that a more promising route to what has previously been studied is to boost the denominator in Eq.~(\ref{eq:RKeqn}),  by allowing the leptoquark to couple to electrons.\footnote{A model independent discussion of couplings to electrons and muons using effective field theory has been performed in Ref.~\cite{Hiller:2014ula}.} As the relevant operator is generated at tree level, the required couplings are quite small. The deviations in the LFU ratios $R_{D^{(*)}}$ can be explained at the same time by introducing a coupling of the $R_2$ leptoquark to $\tau$ leptons.
A mild tension with the theoretically inferred constraint on BR$(B_c\to\tau\nu)$~\cite{Akeroyd:2017mhr} can be resolved by the introduction of the $S_3$ leptoquark, which can be motivated within a radiative neutrino mass model.

The structure of this paper is as follows. In section \ref{Sec:EFT} we
perform an effective field theory (EFT) analysis of the $R_2$ leptoquark. We then explain the most relevant constraints
in section \ref{Sec:constraints} and show that $R_2$ can explain $R_{K^{(*)}}$
and $R_{D^{(*)}}$. In section~\ref{Sec:both} we introduce a minimal model for neutrino masses based on the $R_2$ and $S_3$ leptoquarks.
Finally we conclude in section \ref{Sec:conclusion}.

\mathversion{bold}
\section{Effective field theory analysis for the $R_2$ leptoquark}\label{Sec:EFT}
\mathversion{normal}
The $R_2\sim(3,2,7/6)$ leptoquark is an electroweak doublet and couples to both left-handed and right-handed SM quarks and leptons. Its Yukawa couplings with SM fermions are 
\begin{eqnarray}
{\cal L} _{\rm R2} &=& 
-\left(Y_{2}\right)_{ab} \bar u_a R_2^\alpha \epsilon_{\alpha\beta} P_L L^\beta_b - \left(Y_4\right)_{ab} \bar e_a  R_2^\dagger P_L Q_b + h.c.  \;.
\end{eqnarray}
We work in the basis, where the flavor eigenstates of down-type quarks and charged leptons coincide with their mass eigenstates.
In particular the component of $R_2$ with electric charge $2/3$ couples right-handed charged leptons to left-handed down-type quarks and right-handed up-type quarks to neutrinos and thus contributes to both $b\to s$ and the $b\to c$ processes.

\begin{figure}[tb]
\centering
\includegraphics[width=0.45\textwidth]{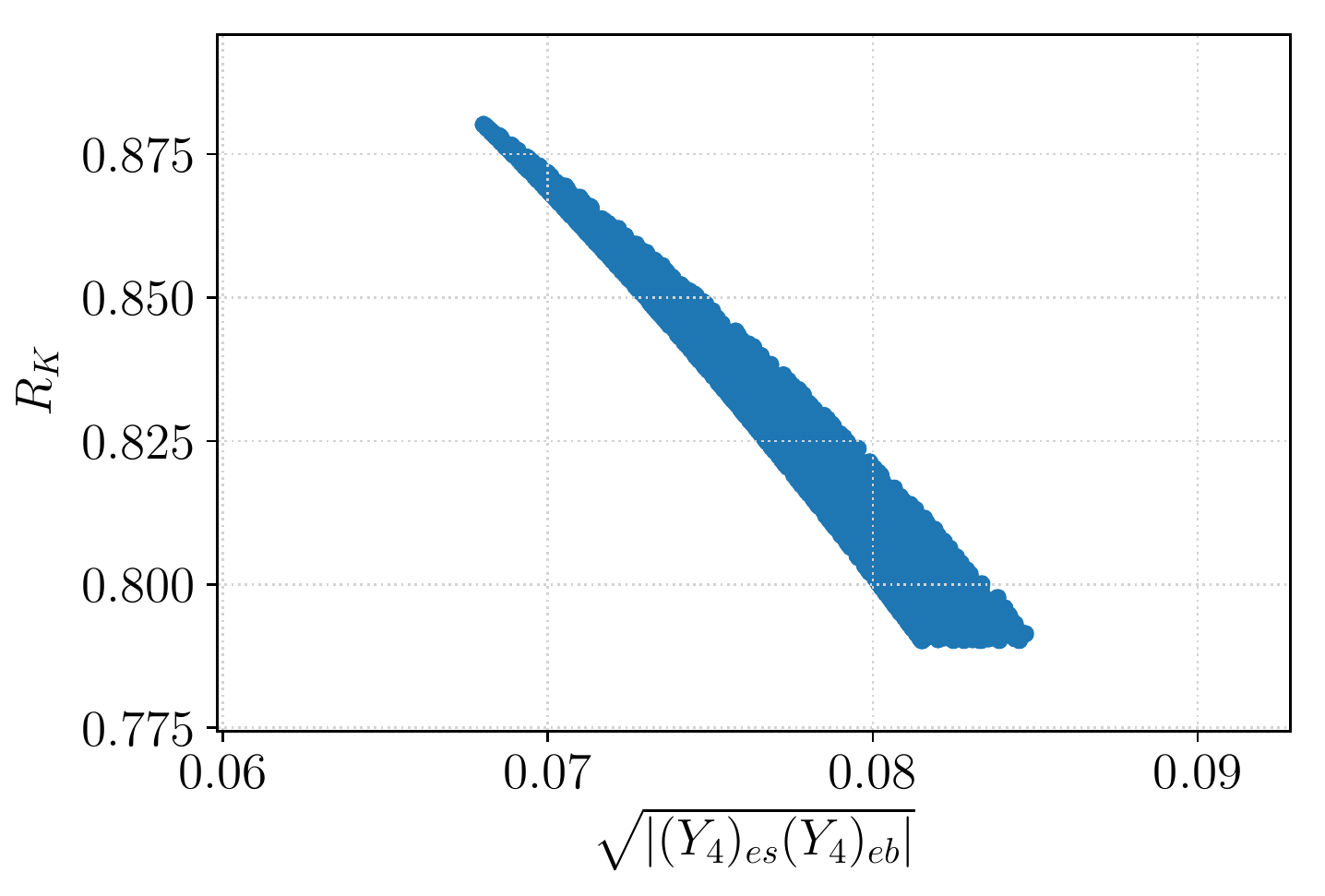} 
\includegraphics[width=0.45\textwidth]{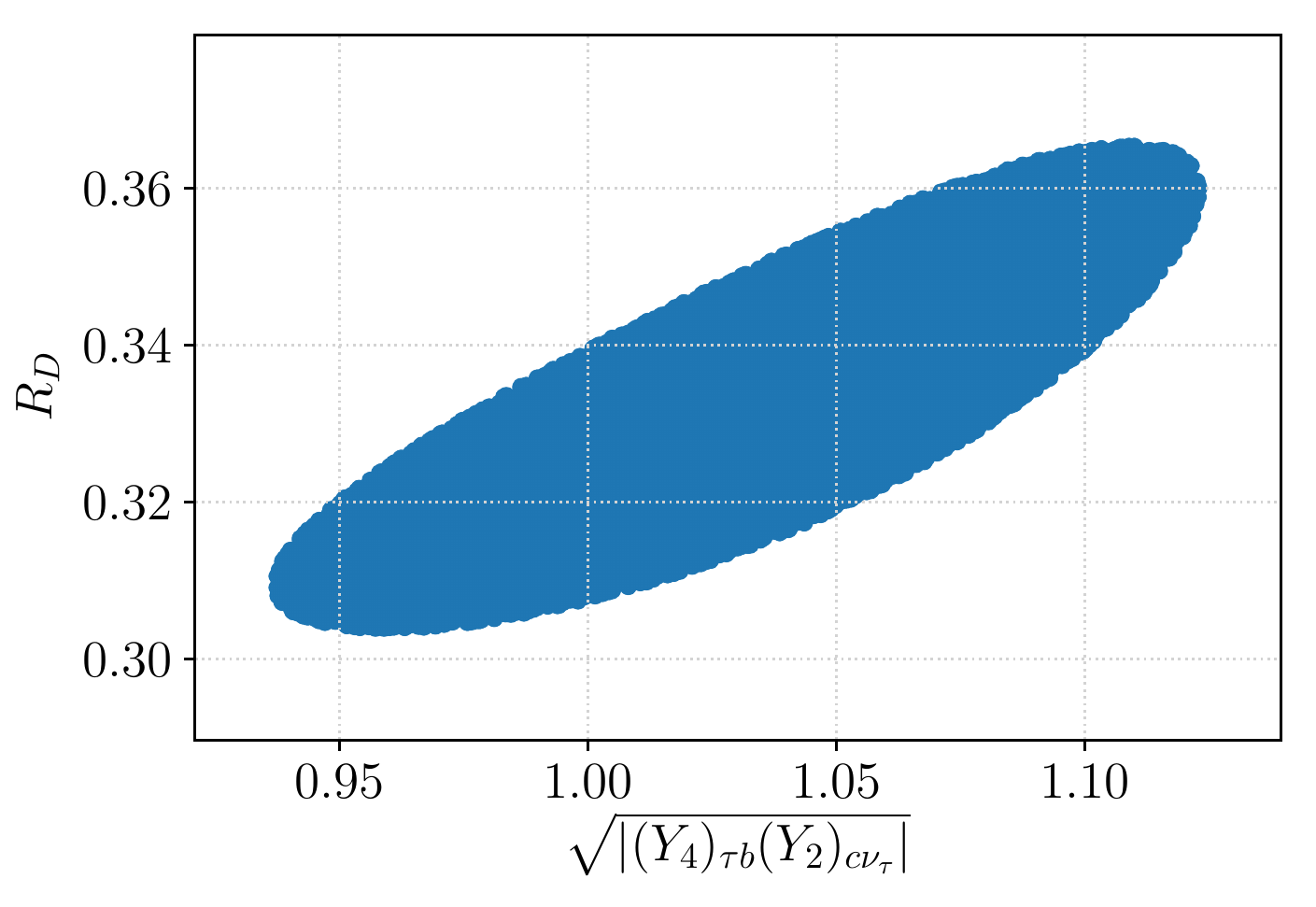} 
    \caption{Dependence of $R_{D^{(*)}}$ and $R_{K^{(*)}}$ on the relevant Yukawa couplings for fixed leptoquark mass $m_{R_2}=1$ TeV.}
    \label{fig:separate}
\end{figure}

For energies below the mass of the leptoquark, it is convenient to write an effective Lagrangian to capture the relevant contributions beyond the SM. Using the Warsaw basis~\cite{Grzadkowski:2010es} of the SM effective field theory (SMEFT), the relevant terms in the effective Lagrangian are
\begin{align}
    \mathcal{L} &= 
    C_{abcd}^{qe} \left( \bar{Q}_a \gamma _\mu Q_b \right) \left( \bar{e}_c \gamma ^\mu e_d \right)% \\ &&
    +C_{abcd}^{lu} \left( \bar{L}_a \gamma _\mu L_b \right) \left( \bar{u}_c \gamma ^\mu u_d \right) \\\nonumber &
    + C_{abcd} ^{lequ1} \left( \bar{L}^j _a e_b  \right) \epsilon _{jk} \left( \bar{Q}_c^k u_d \right) + C_{abcd}^{lequ3} \left( \bar{L}_a^j \sigma _{\mu \nu} e_b \right) \epsilon _{jk} \left( Q_c^k \sigma ^{\mu \nu} u_d \right) \ ,
\end{align}
where $\sigma _{\mu \nu} = \frac{i}{2} \left[ \gamma _\mu , \gamma _\nu \right]$ with Wilson coefficients
\begin{eqnarray}
C_{bdca}^{qe}(m_{R_2})&=& - \frac{\left(Y_{4} \right)^*_{ab}\left(Y_{4}\right)_{cd}}{2m_{R_2}^2} 
\qquad\qquad C_{dbac}^{lu}(m_{R_2}) = - \frac{\left(Y_{2}\right)_{ab}\left( Y_{2}\right)^*_{cd}}{2m_{R_2}^2} \\
C_{bcda}^{lequ1}(m_{R_2}) &=& 4 C_{bcda}^{lequ3 } (m_{R_2})= -\frac{\left(Y_{2} \right)_{ab}\left(Y_{4}\right)_{cd}}{2m_{R_2}^2} \ ,
\end{eqnarray}
which are defined at the renormalization scale $\mu=m_{R_2}$, the mass of leptoquark $R_2$.
The vector Wilson coefficient $C^{qe}_{sbee}$ contributes to $b\to s ee$ and thus modifies the LFU ratios $R_{K^{(*)}}$. This is illustrated in the left panel of Fig.~\ref{fig:separate}. The blue-shaded region indicates the $1-\sigma$-allowed region for $R_{K^{(*)}}$. For a fixed leptoquark mass $m_{R_2}=1$ TeV, the LFU ratio $R_K$ decreases when increasing the magnitude of the Yukawa couplings $|(Y_{4})_{es} (Y_4)_{eb}|$, thus increasing the magnitude of the Wilson coefficient $C^{qe}_{sbee}$.

Similarly the scalar and tensor Wilson coefficients $C^{lequ1,3}_{\nu\tau bc}$ contribute to $b\to c\tau \nu$ and thus modify the LFU ratios $R_{D^{(*)}}$. As the final state neutrino is not measured, there is a contribution from all three flavors. 
The coupling to $\nu_{e,\mu}$ is accompanied by couplings to $e$ and $\mu$ respectively and hence there are additional constraints from lepton flavor violating processes. In order to avoid these additional constraints, we only consider couplings to $\nu_\tau$. The dependence of $R_D$ to the magnitude of the Yukawa couplings $|(Y_2)_{c\nu_\tau} (Y_4)_{\tau b}|$ is illustrated in the right panel of Fig.~\ref{fig:separate}. The blue-shaded region indicates the $1-\sigma$-allowed region for $R_{D^{(*)}}$. The Yukawa couplings $|(Y_2)_{c\nu_\tau}|$ and $|(Y_4)_{\tau b}|$ are generally of order 1 with $\sqrt{|(Y_2)_{c\nu_\tau} (Y_4)_{\tau b}|}\sim 1$ and thus typically larger than the Yukawa couplings required to explain $R_{K^{(*)}}$. As there is generally operator mixing, when evolving the Wilson coefficients from the scale of the leptoquark to the scale of the $b-$quark, the large Wilson coefficients $C_{\nu_\tau\tau bc}^{lequ1,3}$ typically modify the result for $R_{K^{(*)}}$ and thus the interesting parameter range for the Yukawa couplings $(Y_4)_{es}$ and $(Y_4)_{eb}$ differs, when attempting to explain both $R_{K^{(*)}}$ and $R_{D^{(*)}}$ simultaneously. 

A minimal set of Yukawa couplings to accommodate a simultaneous solution to $R_{K^{(*)}}$ and $R_{D^{(*)}}$ is
\begin{equation}
    Y_2= \left( \begin{array}{ccc}
         0 & 0 & 0  \\
         0 & 0 & (Y_2)_{c \nu _\tau }\\ 
         0 & 0 & 0  \\
    \end{array} \right) , \qquad\qquad Y_4 = \left( \begin{array}{ccc}
         0 &  (Y_4)_{es} & (Y_4)_{eb}  \\
         0 &0 & 0 \\
         0 &0 & (Y_4)_{ \tau b}
    \end{array} \right) \ ,
\end{equation}
which we focus on in the following.

Before discussing the phenomenology of the $R_2$ leptoquark we briefly make a connection to the operators in the commonly used operator basis in $B$-physics. We limit our discussion to the operators induced after integrating out the $R_2$ leptoquark. In the weak effective theory, after integrating out the Higgs, $Z$- and $W$-bosons and the top quark, the relevant operators in the effective Lagrangians governing $b\to s ll$ and $b\to c\ell\nu$ decays are
\begin{align}
	\mathcal{L}_{sb\ell\ell} &= \frac{4 G_F}{\sqrt{2}} V_{tb} V_{ts}^* \frac{\alpha_{em}}{4\pi} \sum_{\ell} \left[ C_9^\ell (\bar s\gamma_\mu P_L b ) (\bar \ell \gamma^\mu \ell)  + C_{10}^\ell(\bar s\gamma_\mu P_L b ) (\bar \ell \gamma^\mu \gamma_5\ell) \right] 
	\\\nonumber
 \mathcal{L}_{cb\ell\nu} &= -\frac{4 G_F}{\sqrt{2}} V_{cb} \sum_{i,j} \Big[
      C_V^{ij}(\bar{c} \gamma^\mu P_L b)(\bar{\ell}_i \gamma_\mu P_L  \nu_{j}) 
      +  C^{ij}_S (\bar{c}P_L b)(\bar{\ell}_i P_L\nu_{j})  + C^{ij}_T (\bar{c} \sigma^{\mu \nu} P_L b)
      (\bar{\ell}_i \sigma_{\mu \nu} P_L \nu_{j})\Big] \;,
\end{align}
respectively, with the CKM mixing matrix elements $V_{ij}$. The Wilson coefficients in weak effective theory are related to the ones in SMEFT by 
\begin{align}
C_9^e &= C_{10}^e = \frac{\pi\; C^{qe}_{bsee} }{\sqrt{2}V_{tb}V_{ts}^* G_F \alpha_{\rm em}} & 
C_S^{\tau\nu_\tau} &= 4 C_T^{\tau\nu_\tau} = -\frac{ C_{\nu_\tau\tau bc}^{lequ1}}{2\sqrt{2} V_{cb} G_F}\;.
\end{align}
In our numerical analysis we use the flavio package \cite{Straub:2018kue} for the renormalization group evolution of the Wilson coefficients and the calculation of most processes. We vary the magnitude of the four Yukawa couplings over the range consistent with perturbativity and the explanation of the $R_{D^{(*)}}$ and $R_{K^{(*)}}$ anomalies at $1-\sigma$ and their phases over the whole allowed range $[0,2\pi]$ while fixing the mass $m_{R_2}=1$ TeV.

\section{Experimental constraints and the viable parameter space}\label{Sec:constraints}

In this section we first summarize the most significant constraints on the couplings of the $R_2$ leptoquark in Sec.~\ref{sec:constraint-list}, followed  by a discussion of the viable parameter space in Sec.~\ref{sec:parameter-space}.

\begin{figure}[tb]
    \centering
     \includegraphics[width=0.45\textwidth]{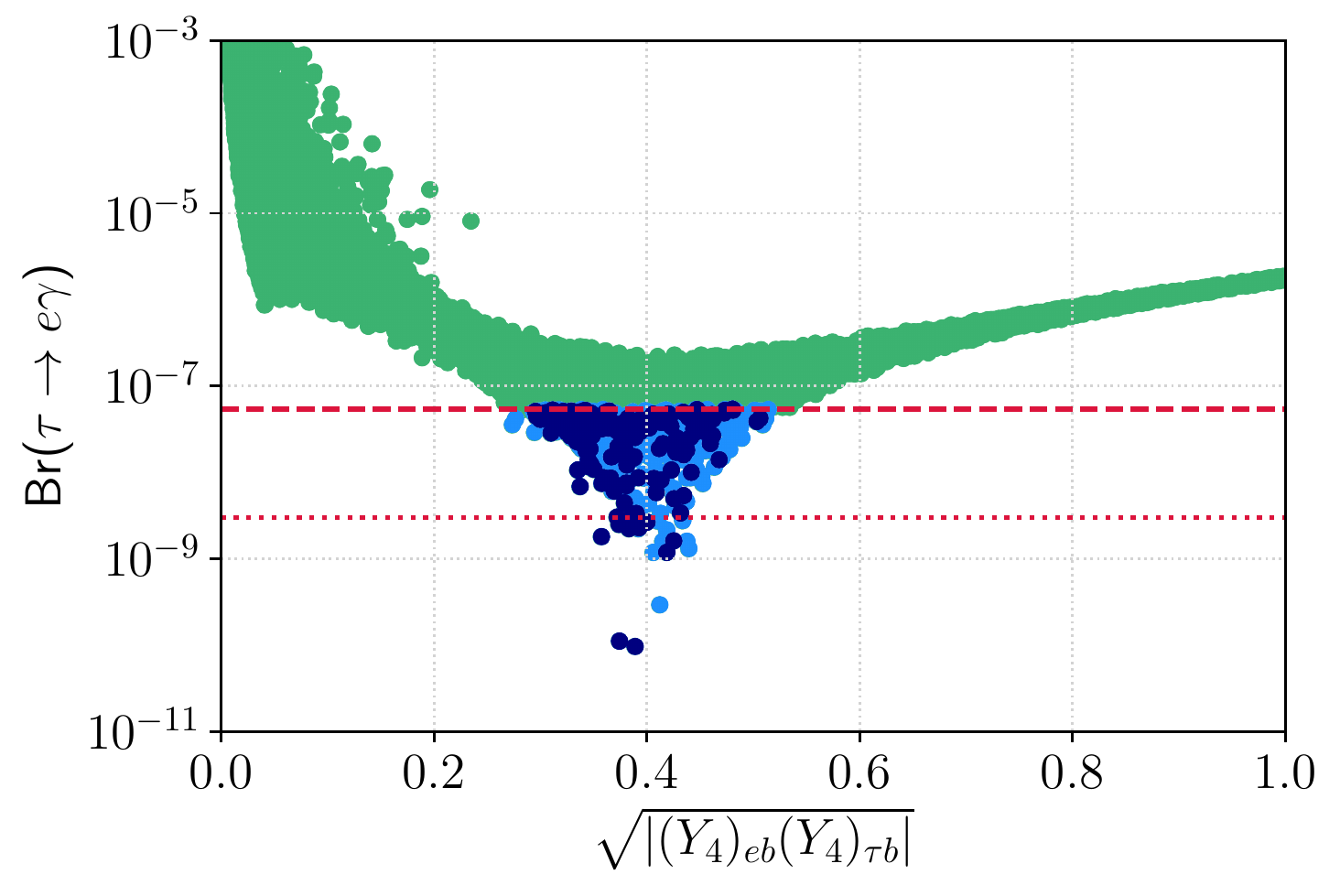} 
\includegraphics[width=0.45\textwidth]{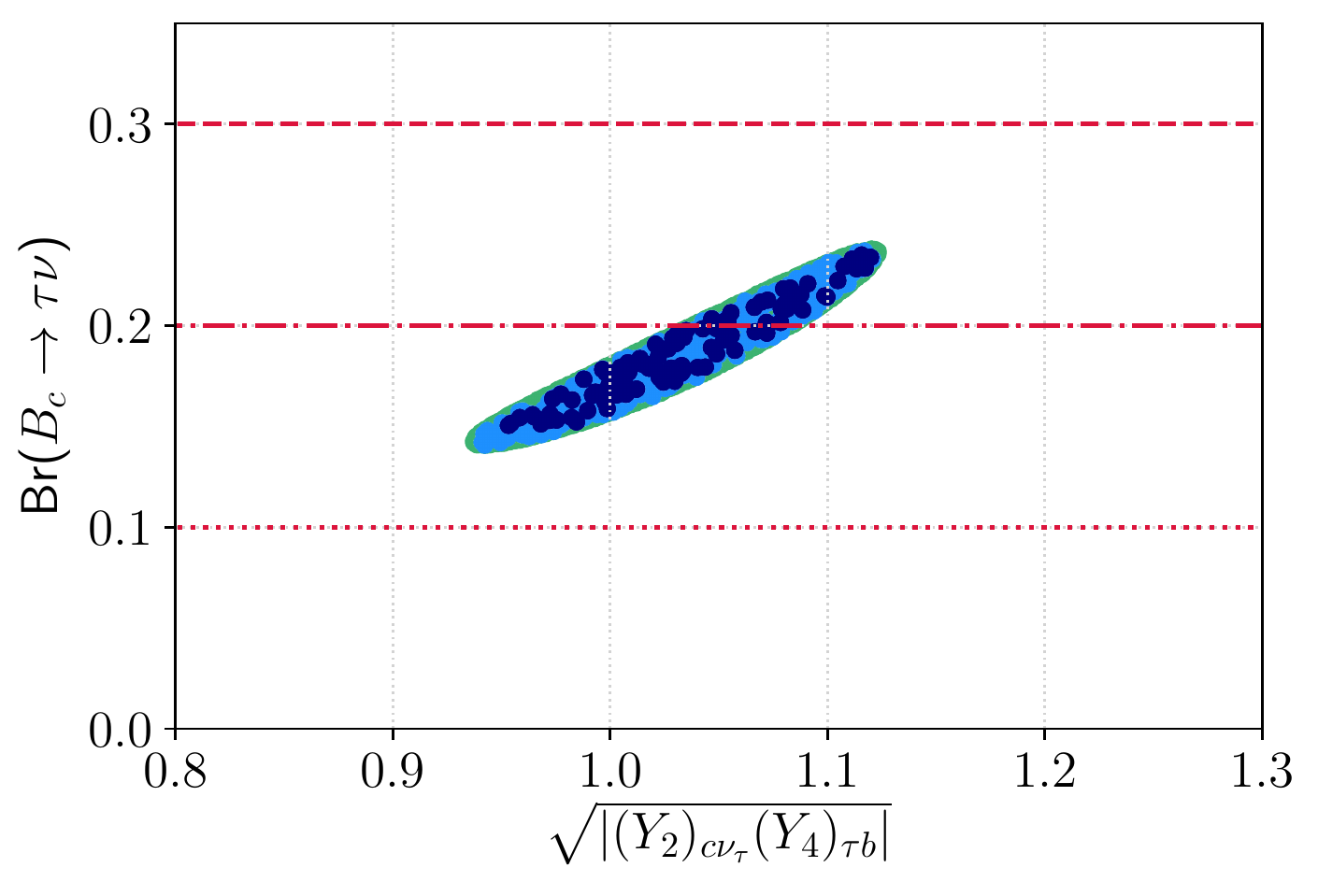} \\
   \includegraphics[width=0.45\textwidth]{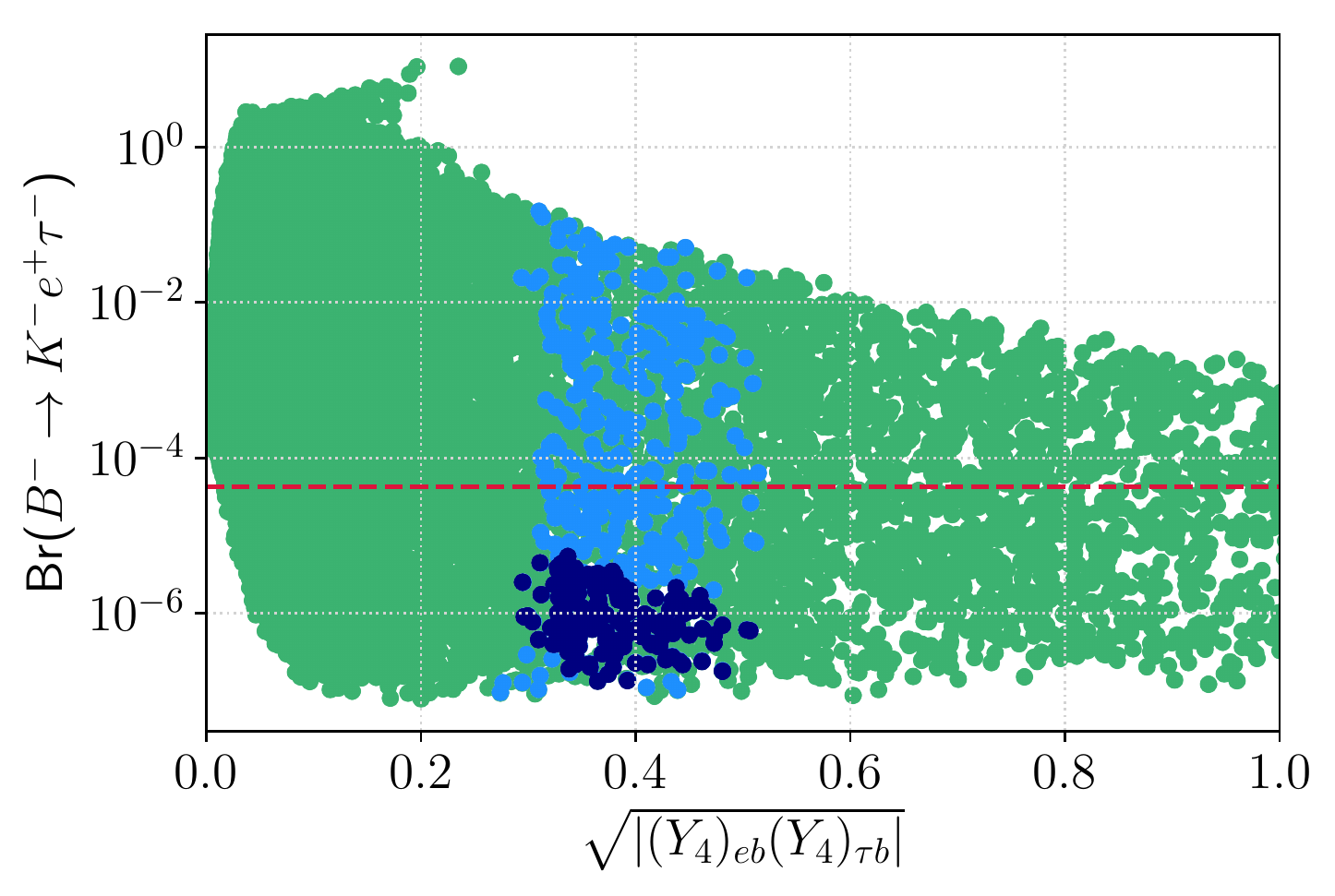}
    \includegraphics[width=0.45\textwidth]{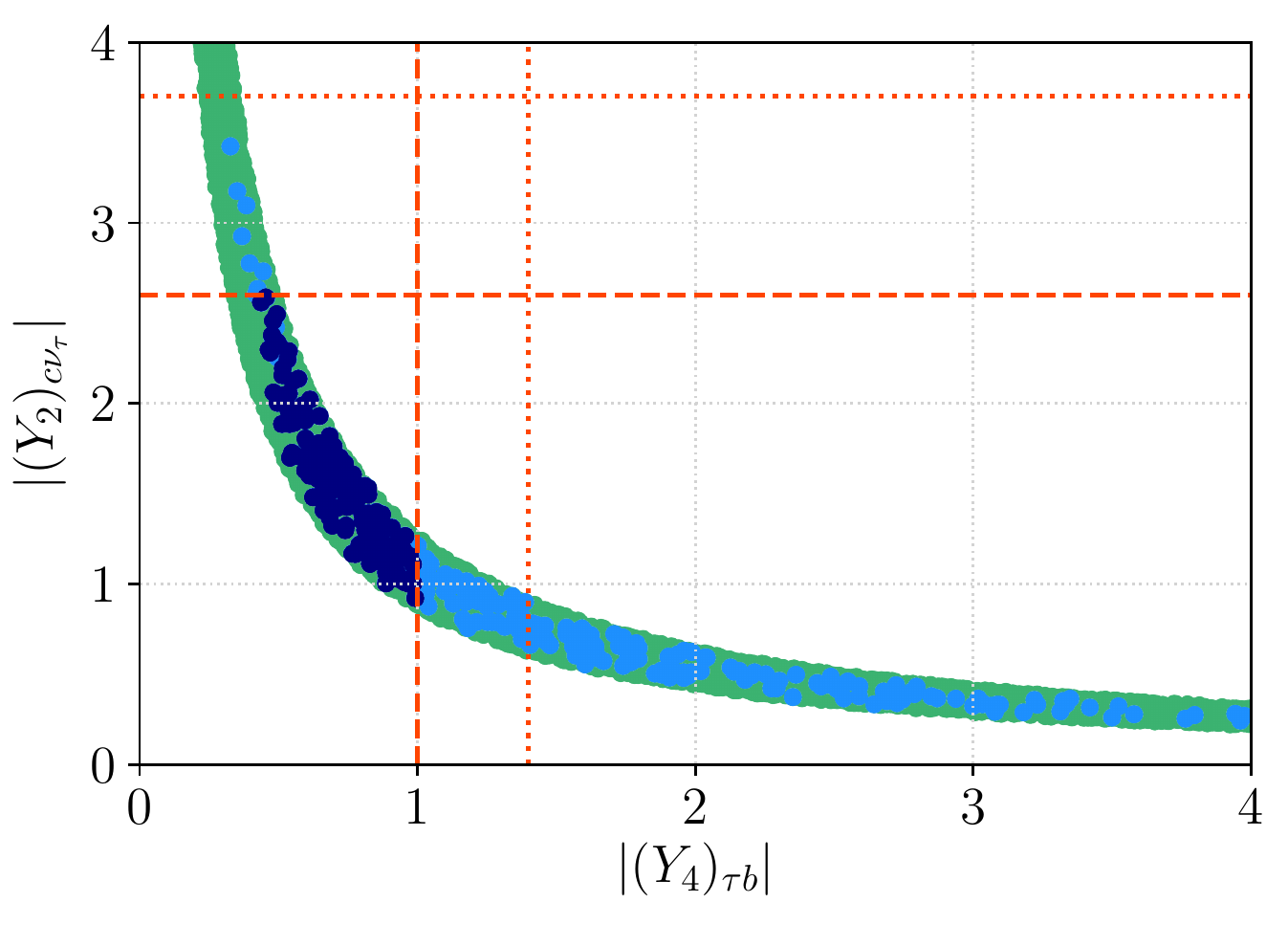} 
    \caption{Points that simultaneously explain $R_{D^{(*)}}$ and
    $R_{K^{(*)}}$ at $1-\sigma$ with observables BR($\tau^- \to e^- \gamma$),
    BR($B_c\to \tau \nu$), BR($B^-\to K^-e^+\tau^-$) shown in the top left, top
    right, and bottom left panels respectively. The bottom right panel shows the
    relative size of the largest Yukawa couplings. The orange dashed (dotted)
    gridlines indicate the contributions to the $Z\tau\tau$ coupling at the
    level of the $1-\sigma$ ($2-\sigma$) experimental error. All points explain
    $R_{K^{(*)}}$ and $R_{D^{(*)}}$ at the $1-\sigma$ level. Dark blue points
    satisfy all constraints. Light blue points satisfy strict limits on
    BR($\tau^- \to e^- \gamma$) but are excluded by other constraints. In the
    top right we show the three most stringent bounds on BR($B_c\to \tau\nu$)
    inferred from different theoretical
    considerations~\cite{Alonso:2016oyd,Akeroyd:2017mhr,AkeroydPrivateCommunication}
    and note that only the most stringent one rules out the scenario.}
    \label{fig:constraints}
\end{figure}

\subsection{Constraints}\label{sec:constraint-list}

We show the impact of the most relevant constraints on the parameter space satisfying a $1-\sigma$ simultaneous solution for $R_{K^{(*)}}$ and $R_{D^{(*)}}$ in Fig.~\ref{fig:constraints}. The four most stringent constraints are posited by the decays $\tau\to e\gamma$, $B^+\to K^+\tau^+e^-$, $Z\to \tau\tau$ and $B_c\to\tau\nu$.

\mathversion{bold}
\subsection*{$\tau^-\to e^- \gamma$}
\mathversion{normal}

The radiative lepton-flavor-violating decay $\tau^-\to e^-\gamma$ occurs at loop level. Its branching ratio takes the form~\cite{Lavoura:2003xp}
\begin{equation}
    \frac{\mathrm{BR}(\tau ^- \to e^- \gamma)}{\mathrm{BR}(\tau ^- \to e^- \nu_\tau \bar\nu_e)} \simeq \frac{27 \alpha_{em}}{256\pi G_F^2 m_{R_2}^4} 
    \left| (Y_4^* Y_4^T)_{\tau e} - \frac43 \sum_{q=u,c,t} (Y_2)_{q\tau} (Y_4V^\dagger)_{eq}  \frac{ m_q}{m_\tau} \left(1-\ln\frac{m_q^2}{m_{R_2}^2}\right)\right|^2
    \label{eq:tautoegamma} 
\end{equation}
in the limit of vanishing final state electron mass and to leading order in the quark masses in the loop. The branching ratio for the SM purely leptonic $\tau$ decay is BR($\tau\to e\nu_\tau\bar\nu_e)=0.178$. In the numerical scan, we use the exact expression and impose the current limit on the branching ratio BR$(\tau ^- \to e^- \gamma )<5.4 \times 10^{-8}$ obtained by HFLAV~\cite{Amhis:2016xyh}. The HFLAV limit is less stringent than the limit quoted in the PDG, because it combines the BABAR result \cite{Aubert:2009ag} with the less stringent Belle result~\cite{Hayasaka:2007vc} while the PDG~\cite{Tanabashi:2018oca} relies only on the former. The combined limit is less aggressive as Belle saw a small excess of this process (see table 319 in Ref.~\cite{Amhis:2016xyh}). Irrespective whether the Belle result is included or not, the simultaneous explanation of both $R_{K^{(*)}}$ and $R_{D^{(*)}}$ is viable.

In the top left panel of Fig.~\ref{fig:constraints} we show the branching ratio vs $(|(Y_4)_{eb}(Y_4)_{\tau b}|)^{1/2}$. For large couplings $|(Y_4)_{eb}(Y_4)_{\tau b}|$, the branching ratio is dominated by the first term and thus increases for increasing Yukawa couplings. For small $(Y_4)_{\tau b}$, the Yukawa coupling $(Y_2)_{c\nu_\tau}$ becomes large in order to explain $R_{D^{(*)}}$ as shown in the bottom right plot and thus the second term in Eq.~\eqref{eq:tautoegamma} dominates, which explains the increasing branching ratio for small $|(Y_4)_{eb}(Y_4)_{\tau b}|$. The Belle II experiment~\cite{Kou:2018nap} is expected to improve the sensitivity to $\tau\to e\gamma$ by more than one order of magnitude to $3\times 10^{-9}$ (indicated by a dotted red line) and thus probe a large part of the remaining parameter space.

\mathversion{bold}
\subsection*{$B^+\to K^+\tau^+ e^-$}
\mathversion{normal}
Another constraint on the $\tau-e$ flavor violating processes originates from the semileptonic lepton flavor violating $B$ decay $B^+\to K^+\tau^+ e^-$. Its branching ratio satisfies BR$(B^+\to K^+ \tau^+ e^-)<1.5\times10^{-5}$~\cite{Tanabashi:2018oca}. The $R_2$ leptoquark induces the vector operator
\begin{equation}
    C^{qe}_{sb\tau e} = - \frac{(Y_4^*)_{es} (Y_4)_{\tau b}}{2m_{R_2}^2}\;,
\end{equation}
which contributes to $B^+\to K^+ \tau^+e^-$ and thus constrains the simultaneous explanation of $R_{K^{(*)}}$ and $R_{D^{(*)}}$.
As we demonstrate in the bottom left panel of Fig.~\ref{fig:constraints}, it provides a moderate constraint on the parameter space that simultaneously explains $R_{K^{(*)}}$ and $R_{D^{(*)}}$. The region excluded by $B^+\to K^+\tau^+e^-$ is also excluded by $\tau\to e\gamma$. 

\mathversion{bold}
\subsection*{$Z$ decays}
\mathversion{normal}
The $R_2$ leptoquark also contributes to several $Z$-boson decay processes. In particular, its contribution to $Z\to \tau\tau$ is significant due to the large couplings to $\tau$ leptons. Approximate expressions for the left-handed and right-handed couplings of the $Z$-boson to $\tau$ leptons
\begin{align}
    \mathrm{Re}(\delta g^\tau_L) &\simeq \frac{|(Y_4)_{\tau b}|^2}{16\pi^2} \left\{  -\frac32 x_t \left[1+\ln x_t\right] +x_Z\left[\frac{23}{12}+\left(\frac{128}{9} +8 \ln x_t -\frac13\ln x_Z\right)\sin^2\theta_W \right]  \right\}
    \\
    \mathrm{Re}(\delta g^\tau_R) &\simeq \frac{|(Y_{2})_{c\nu_\tau}|^2}{16\pi^2}  x_Z \left[\frac{1}{12}-\frac12 \ln x_Z +\left(\frac{1}{18}+\frac23 \ln x_Z\right)\sin^2\theta_W\right]
\end{align}
in terms of the Weinberg angle $\theta_W$ and the ratios $x_t=(m_t/m_{R_2})^2$ and $x_Z=(m_Z/m_{R_2})^2$ are obtained by expanding the expressions given in Ref.~\cite{Arnan:2019olv} to leading order in $x_t$, $x_Z$ and the quark mixing angles by taking $V_{tb}\simeq 1$.

The LEP experiments measured the $Z$-boson couplings precisely~\cite{ALEPH:2005ab} with $1-\sigma$ uncertainties of $|\mathrm{Re}(\delta g_L^{\tau})| < 5.8 \times 10^{-4}$ for couplings to left-handed $\tau$ leptons and $|\mathrm{Re}(\delta g_R^{\tau})| \leq 6.2 \times 10^{-4}$ for right-handed $\tau$ leptons. This translates to a constraint on the magnitude of the Yukawa couplings $|(Y_4)_{\tau b}|$ and $|(Y_2)_{c\nu_\tau}|$ of $|\left( Y_4 \right)_{\tau b}| \lesssim 1.0 (1.4)$ and $|\left(Y_2\right)_{c\nu_\tau}|\lesssim 2.6 (3.7)$ using $1-\sigma$ ($2-\sigma $) experimental uncertainties respectively. This is indicated in the bottom right panel of Fig.~\ref{fig:constraints} as orange dashed (dotted) lines. The dark blue points in the numerical scan do not lead to any correction larger than the $1-\sigma$ experimental uncertainties.
 In reality, a full global fit to all electroweak observables is needed to impose a reliable constraint, and it is probable that significantly larger deviations to effective Z couplings can be accommodated. We leave such a work to the future and comment here that even our pessimistic approach does not rule out our model.

\mathversion{bold}
\subsection*{$B_c\to \tau\nu$}
\mathversion{normal}
The $R_2$ leptoquark contributes to $B_c\to\tau\nu$ via the same couplings relevant to $R_{D^{(*)}}$, since the same scalar operator contributes to both $R_{D^{(*)}}$ and $B_c\to \tau\nu$. In the top right panel of Fig.~\ref{fig:constraints} we show the prediction for BR$(B_c\to\tau\nu)$. We find branching ratios between 15\% and 23\% for the region of parameter space which explains both $R_{D^{(*)}}$ and $R_{K^{(*)}}$  at $1-\sigma$. Thus limits on this process pose a direct constraint on the explanation of $R_{D^{(*)}}$. 

Several groups inferred limits on BR$(B_c \to \tau \nu )<[0.1,0.6]$~\cite{Li:2016vvp,Alonso:2016oyd,Akeroyd:2017mhr,Blanke:2018yud,Bardhan:2019ljo} 
via different theoretical arguments. In the top right panel of
Fig.~\ref{fig:constraints} we indicate the three most stringent theoretically
argued upper limits: 10\%~\cite{Akeroyd:2017mhr},
20\%~\cite{AkeroydPrivateCommunication} and 30\%~\cite{Alonso:2016oyd}, which
are indicated by a dotted, dash-dotted, and dashed red line, respectively. In
particular, Ref.~\cite{Akeroyd:2017mhr} found that the branching ratio can be
at most 10\%, which is in tension with the viable parameter space of the $R_2$
leptoquark explanation of $R_{D^{(*)}}$. However, there is some controversy
over this constraint that relies on the probability that a bottom quark
hadronizes with a charmed quark that has not yet been measured at the LHC.
There is currently an ongoing discussion on how the probability at the LHC can
be inferred from LEP measurements and Monte Carlo simulations, where the
authors of the first paper~\cite{Akeroyd:2017mhr,AkeroydPrivateCommunication}
advocate values in the range $10-20\%$ compared to some recent papers that
advocate liberal constraints in the range
$39-60\%$~\cite{Blanke:2018yud,Bardhan:2019ljo}. The large range
of values comes from the fact that the limit is quite sensitive to the ratio of
hadronization probabilities $f_c/f_u$, that is the probability of hadronizing
with a charm or up quark, respectively. Fig. 5 in Ref. \cite{Bardhan:2019ljo} in
particular shows this dependency where a bound of $10 \%$ corresponds to
$f_c/f_u \sim 4\times 10^{-3}$ and the most liberal bound of $60\%$ corresponds
to a ratio that is a factor of 5 smaller, $f_c/f_u \sim 8\times 10^{-4} $. Reference~\cite{Bardhan:2019ljo} use Pythia8 \cite{Sjostrand:2006za,Sjostrand:2014zea},
which relies on experimental data to tune hadronization, to estimate the 
hadronization fraction and derive a fairly liberal lower bound of $39\%$.
However, they and others \cite{AkeroydPrivateCommunication} express skepticism
of the true lower bound.  We remain agnostic with regards to this constraint
and show our prediction for this branching ratio in the top right panel of
Fig.~\ref{fig:constraints}.

Even the most stringent constraint of 10\% may be avoided by extending the model with a $S_3$ leptoquark, as we discuss in Sec.~\ref{Sec:both}.

\subsection*{Other constraints}
Apart from the discussed constraints we also studied possible constraints from several other processes and we briefly summarize the results.
The limits obtained from lepton-flavor-violating semileptonic $\tau$ decays, $\tau\to e P$ with a pseudoscalar meson $P=K,\pi$, are always substantially weaker than the limit from $\tau\to e\gamma$ and thus we do not report them here. Furthermore, we considered leptonic meson decays, in particular $B_s\to ee$, $D_0\to ee$ and $D_s\to e\nu$, using flavio and as expected neither of them provides a relevant constraint. As the couplings required for an explanation of $R_{K^{(*)}}$ are small, the contribution to $B_s\to ee$ is suppressed. The dominant contribution to $D_s\to e\nu$ is controlled by $(Y_2)_{c\nu_\tau} (Y_4)_{es}$. While $(Y_4)_{es}$ is small, $(Y_2)_{c\nu_\tau}$ is constrained by its contribution to $\tau\to e\gamma$. Moreover, the contribution to the LFU ratios $R_D^{\mu/e} \equiv \Gamma(B\to D \mu\mu)/\Gamma(B\to D ee)$ and $R_{D^*}^{e/\mu} \equiv\Gamma(B\to D^* ee )/\Gamma(B\to D^* \mu\mu)$ are generally small, because the couplings $(Y_4)_{es}$ and $(Y_4)_{eb}$ that are responsible for explaining $R_{K^{(*)}}$ are small. 

Let us turn our attention to $B_s-\bar B_s$ mixing. Matching the full theory with the $R_2$ leptoquark to SMEFT induces an effective four-quark interaction in SMEFT
\begin{equation}
    \mathcal{L} = -\frac{(Y_4^\dagger Y_4)^2_{ij}}{128\pi^2 m_{R_2}^2} (\bar Q_i \gamma_\mu Q_j)   (\bar Q_i\gamma^\mu Q_j)\;.
\end{equation}
In particular, this four-quark interaction induces a new contribution to $B_s-\bar B_s$ mixing which can be parameterized by  
\begin{align}
\mathcal{L} &= C_{VLL}^{bsbs} \;(\bar s \gamma^\mu P_L b) \, (\bar s \gamma_\mu P_L b) 
&
C_{VLL,R_2}^{bsbs} & = -\frac{\left(Y_4 ^{\dagger }  Y_4\right) _{sb}^2}{128\pi^2 m_{R}^2} 
\end{align}
in weak effective theory. 
It interferes with the SM contribution (see e.g.~\cite{Fleischer:2008uj})
\begin{equation}
    C_{VLL,SM}^{bsbs} = \frac{G_F^2 m_W^2}{4\pi^2} (V_{tb}^* V_{ts})^2 S_0(m_t^2/m_W^2)
\end{equation}
where $S_0$ is the Inami-Lim function~\cite{Inami:1980fz}
\begin{equation}
    S_0(x) = \frac{x^3-11x^2+4x}{4(x-1)^2}-\frac{3x^3}{2(x-1)^3}\ln x\;.
\end{equation}
The contribution of $R_2$ to $C_{VLL}^{bsbs}$ can be expressed in terms of the Wilson coefficient $C_9^e$. A simple order of magnitude estimate shows that $C_{VLL,R_2}^{bsbs}$ is several orders of magnitude smaller as the SM contribution for the interesting $R_2$ leptoquark mass range
\begin{align}
\left|\frac{C_{VLL,R_2}^{bsbs}}{C_{VLL,SM}^{bsbs}}\right|
\simeq \left( \frac{\alpha_{em}}{2\pi} \frac{m_{R_2}}{m_W} C_9^e\right)^2\;.
\end{align}
We independently checked the contribution to $B_s-\bar B_s$ mixing using flavio with the same result.

Finally, as the deviation in the ratios $R_K^{(*)}$ originates entirely from a
modification to the branching fraction of $B\to K^{(*)}e^+e^-$ one may wonder
whether this deviation is consistent with measurements of the branching ratio
itself.
Under the given assumptions taking the $1\sigma$ boundaries (central value) the
experimental value implies an increase of the binned differential branching
ratio $\langle \tfrac{d\mathrm{Br} (B^+\to K^+e^+e^-)}{dq^2}\rangle
(1.1\,\mathrm{GeV}^2 < q^2<6.0\,\mathrm{GeV}^2)$ by 9\%-21\% (15\%) and similarly
for $B\to K^* e^+e^-$ by 21\%-36\% (28\%).\\
While the experimental error has been reduced to the 10\%
level~\cite{Aaij:2019wad} from 15\%-30\% in earlier
measurements~\cite{Aaltonen:2011qs,Lees:2012tva,Aaij:2014pli,Aaij:2015nea},
there remains a large theory error due to the uncertainties in the hadronic
matrix
elements~\cite{Bailey:2015nbd,Du:2015tda,Bouchard:2013mia,Khodjamirian:2017fxg,Straub:2018kue}.
For $B^+\to K^+e^+e^-$ the error in the different individual theory
calculations is of the order 16\%-34\%. Moreover, the central values differ: To
give an example, flavio~\cite{Straub:2018kue} predicts $(3.49\pm0.53)\times
10^{-8} \mathrm{GeV}^{-2}$ for $1 \,\mathrm{GeV}^2 < q^2< 6\, \mathrm{GeV}^2$,
while Ref.~\cite{Khodjamirian:2017fxg} obtains
$(4.38^{+0.62}_{-0.57}\pm0.28)\times 10^{-8}\, \mathrm{GeV}^{-2}$ with a
central value that is increased by 25\%.
Using flavio we find for the allowed parameter space after imposing all constraints $[3.99,4.42]\times 10^{-8}\,\mathrm{GeV}^{-2}$ which is consistent with the experimental measurement given the large theory uncertainties in $B\to Ke^+e^-$.\\
A similar argument applies for $B\to K^* e^+ e^-$. Flavio predicts $\langle \tfrac{d\mathrm{Br}(B^0\to K^*e^+ e^-)}{dq^2}\rangle (1.1\,\mathrm{GeV}^2 < q^2 < 6\, \mathrm{GeV}^2)=(4.77\pm 0.71) \times 10^{-8}\,\mathrm{GeV}^{-2}$ which corresponds to a theory uncertainty of 15\%, while Ref.~\cite{Jager:2014rwa} quotes errors between 25\% and 100\% depending on the assumptions on the distribution of nuisance parameters. 
Using flavio we find branching ratios in the range $[6.09,7.02]\times 10^{-8}\,\mathrm{GeV}^{-2}$ after imposing all constraints which corresponds to an increase of 28\%-47\% compared to the SM prediction of flavio and thus larger than the SM uncertainty quoted by flavio, but well below the conservative uncertainty estimate in Ref.~\cite{Jager:2014rwa}.\\
In summary, the treatment of the hadronic effects in the theoretical predictions for the branching ratios $B\to K^{(*)}e^+ e^-$ is still the subject of considerable debate, as we illustrated above by referring to the literature, and thus currently $R_{K^{(*)}}$ can be explained by a correction to the semileptonic decay $B\to K^{(*)}e^+ e^-$.

\subsection{Viable parameter space}\label{sec:parameter-space}
\begin{figure}[tb]
    \centering
    \includegraphics[width=0.45\linewidth]{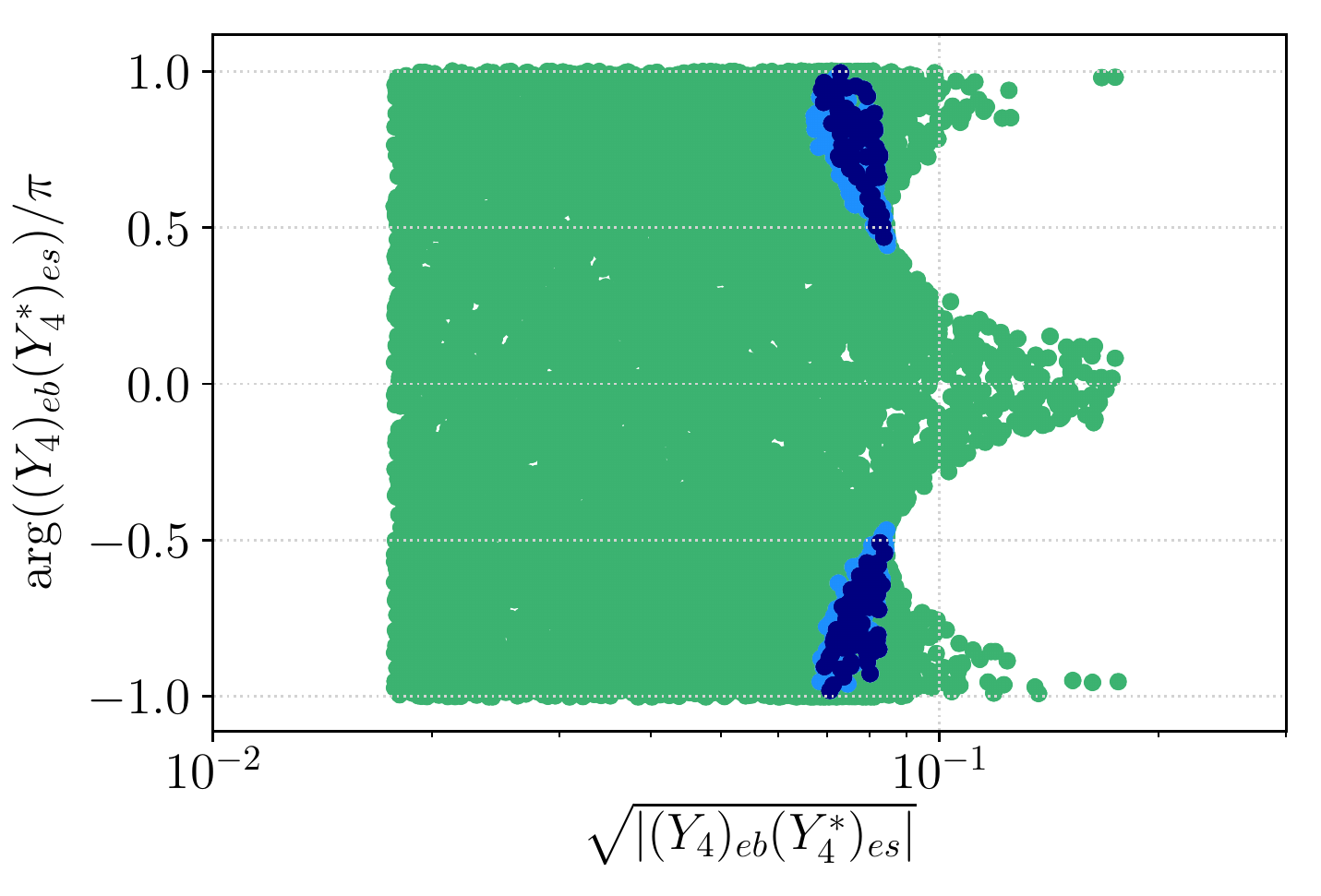}
    \includegraphics[width=0.45\linewidth]{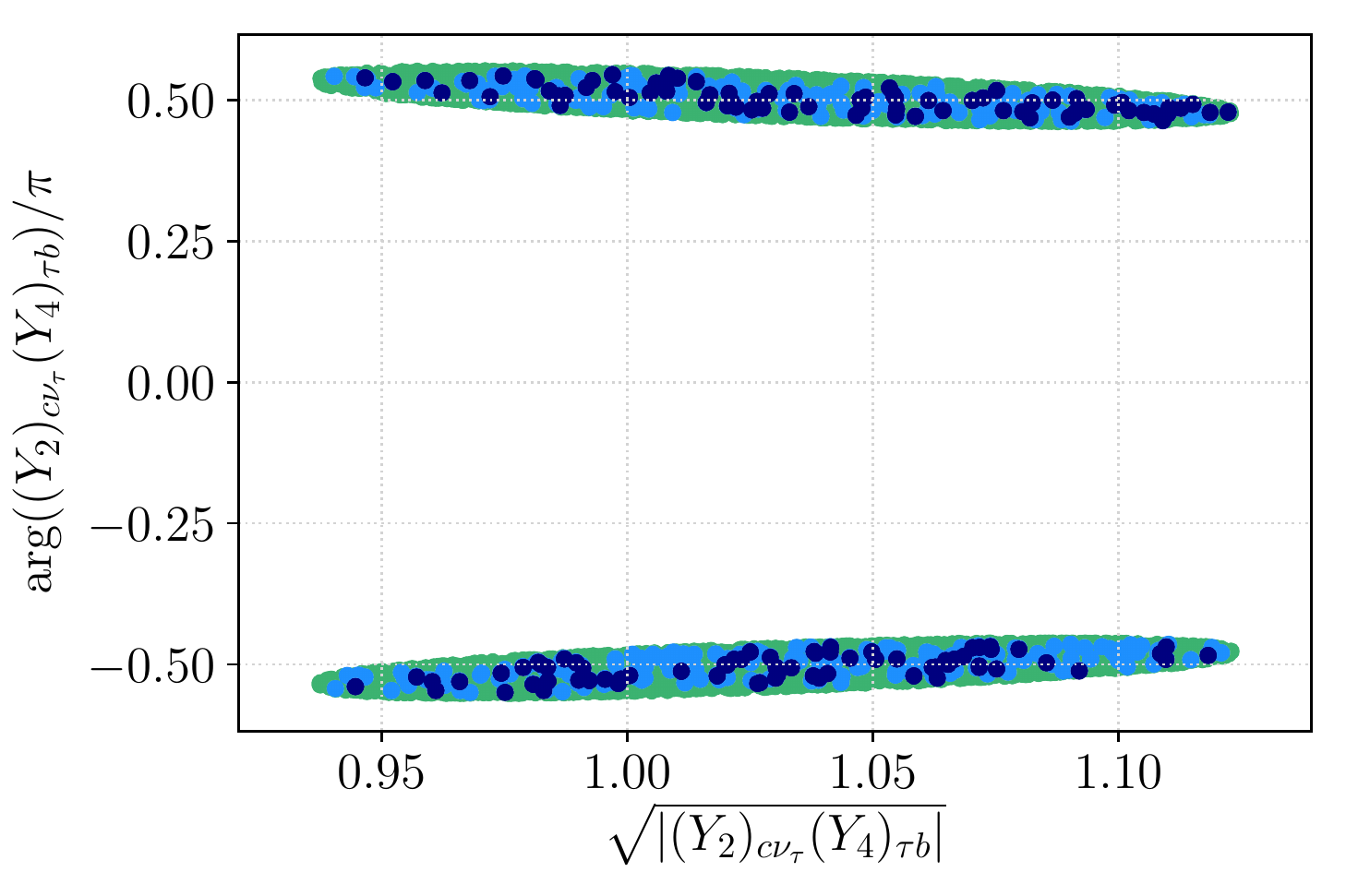}
    
    \includegraphics[width=0.45\linewidth]{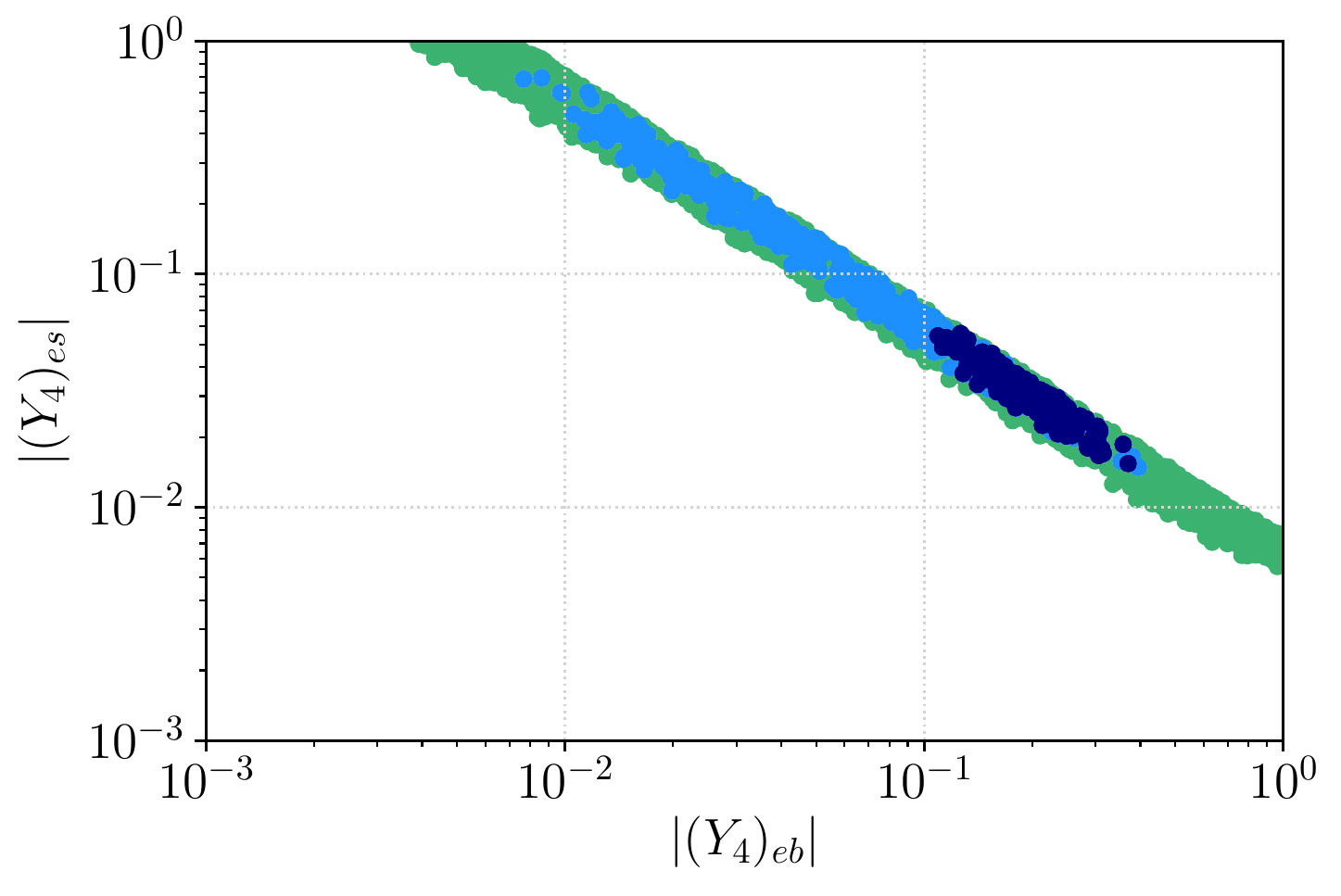}
    
    \caption{Relevant parameter range of the Yukawa couplings. The top left panel shows the phase vs.~the magnitude for $(Y_4)_{eb} (Y_4^*)_{es}$ and the top right panel the phase vs.~the magnitude of $(Y_2)_{c\nu_\tau} (Y_4)_{\tau b}$. In the bottom panel we plot the absolute values of the Yukawa couplings entering $R_{K^{(*)}}$ against each other. All points explain $R_{K^{(*)}}$ and $R_{D^{(*)}}$ at the $1-\sigma$ level. Dark blue points satisfy all experimental constraints. Light blue points satisfy strict limits on $\tau^-\to e^-\gamma$ but are excluded by other constraints.}
    \label{fig:WC}
\end{figure}

We show the viable parameter space in Fig.~\ref{fig:WC} and the bottom right panel of Fig.~\ref{fig:constraints}. 
For a fixed leptoquark mass of $m_{R_2}=1$ TeV, the bottom right panel of  Fig.~\ref{fig:constraints} shows that an aggressive constraint from $Z$ decays restricts two of the Yukawa couplings to the range $|(Y_4)_{\tau b}|\in[0.44,1.0]$ and $|(Y_{2})_{c\nu_\tau}|\in[1.0,2.6]$, respectively. All quoted ranges are approximate and are only intended to give an indication.  The product is also constrained by the need to explain $R_{D^{(*)}}$ at the $1-\sigma$ level to the range $|(Y_2)_{c\nu_\tau} (Y_4)_{\tau b}|\in[0.88,1.3]$.  The product is almost purely imaginary with $\arg((Y_2)_{c\nu_\tau} (Y_4)_{\tau b})\in \pm[0.45,0.54]\pi$,  irrespective of the experimental constraints, which confirms previous findings~\cite{Tanaka:2012nw,Dorsner:2013tla,Sakaki:2013bfa}.

The other two Yukawa couplings are generally smaller with $|(Y_{4})_{eb}|\in[0.11,0.37]$, $|(Y_4)_{es}|\in[0.015,0.055]$. Their product constrained to the narrow range $|(Y_4)_{eb} (Y_4)_{es}|\in[0.0047,0.0070]$ after imposing all experimental constraints. This is shown in the bottom panel of Fig.~\ref{fig:WC}.  For the points that satisfy all experimental constraints, the real part of the product is generally negative, the argument is weakly constrained to the range $\arg((Y_4)_{eb}(Y_4^*)_{es})\in\pm[0.47,1.0]$. This implies that the Wilson coefficient $C_{sbee}^{qe}$ is generally positive. The absolute value of the product is by contrast, confined to a narrow range $\sqrt{|(Y_4)_{eb}(Y_4^*)_{es}|} \in [4.7,7.0]\times 10^{-3}$. The hierarchy $|(Y_{4})_{es}|\ll |(Y_4)_{eb}|$ can be understood as follows. The constraint from $Z$-boson decays constrains the coupling $(Y_4)_{\tau b}\lesssim 1.0$ and thus the coupling $(Y_2)_{c\nu_\tau}$ has to be larger than 1 in order to explain $R_{D^{(*)}}$. This in turn leads to a stronger constraint on $|(Y_2)_{c\nu_\tau}|$ from $\tau\to e\gamma$, because the suppression from the ratio $m_s/m_\tau$ is compensated by the large logarithm, $\ln (m_s^2/m_{R_2}^2)$.

Finally, we obtain a prediction for the branching ratios of the decays $t\to c \ell_i^+\ell_j^-$~\cite{Davidson:2015zza}, $\mathrm{BR}(t\to \ell_i^+\ell_j^- +c) \simeq \tfrac{1.3}{48\pi^2} \left|\epsilon_{ij}^{RL}\right|^2$, where $\epsilon_{ij}^{RL}$ is the Wilson coefficient of the effective operator
\begin{equation}
\mathcal{L} = -2\sqrt{2}G_F \epsilon_{ij}^{RL} [\bar e_i \gamma_\mu P_R e_j] [\bar c\gamma^\mu P_L t]\;.
\end{equation}
The $R_2$ leptoquark especially generates the Wilson coefficients 
\begin{align}
\epsilon_{\tau e}^{RL}    
& = \frac{\left((Y_{4}^*)_{es} V_{cs}^* + (Y_4^*)_{eb} V_{cb}^*\right) (Y_4)_{\tau b} V_{tb} }{4\sqrt{2} G_F m_{R_2}^2}\;,
&
\epsilon_{e\tau}^{RL}
&= \frac{(Y_{4}^*)_{\tau b} V_{cb}^* \left((Y_{4})_{es} V_{ts} + (Y_4)_{eb} V_{tb}\right)}{4\sqrt{2} G_F m_{R_2}^2}\;.
\end{align}
We find that the decays $t\to c \tau^\pm e^{\mp}$ are generally tiny with branching ratios
\begin{align}
    \mathrm{BR}(t\to c\tau^+e^-) & \lesssim 2\times 10^{-9} &
    \mathrm{BR}(t\to c\tau^-e^+) & \lesssim 8\times 10^{-11} 
\end{align}
for the parameter space that explains both $R_{K^{(*)}}$ and $R_{D^{(*)}}$ at the $1-\sigma$ level with a leptoquark mass $m_{R_2}=1$ TeV and thus we do not expect any signal at the LHC experiments.

\mathversion{bold}
\section{Alleviating the (possible) tension with BR$(B_c \to \tau \nu )$ via a neutrino mass model}\label{Sec:both}
\mathversion{normal}

It is well known that the $S_3\sim(\bar 3,3,1/3)$ leptoquark\footnote{The collider phenomenology of an $S_3$ leptoquark has been studied in Ref.~\cite{Alvarez:2018gxs}.} with the Yukawa interaction
\begin{equation}
\mathcal{L}_{S3}= 
	- \left(Y_{S}\right)_{ab} \overline{Q_{ai}^c} P_L L_{bk} S_{3\,jl} \epsilon^{ij}\epsilon^{kl} + \mathrm{h.c.} \;
\end{equation}
and mass $m_S$ contributes to the Wilson coefficients 
\begin{align}
    C_{dbca}^{lq1}&=\frac{3 (Y_S)_{ab}(Y_S^*)_{cd}}{8m_S^2}&
    C_{dbca}^{lq3}&=\frac{(Y_S)_{ab}(Y_S^*)_{cd}}{8m_S^2}&
\end{align}
of the vector operators~\cite{Dorsner:2017ufx,Chen:2017hir,Das:2016vkr}
\begin{equation}
\mathcal{L} = 
C_{abcd}^{lq1} (\bar L_a\gamma_\mu  L_b)(\bar Q_c\gamma^\mu Q_d)
+C_{abcd}^{lq3} (\bar L_a\gamma_\mu \tau^IL_b)(\bar Q_c\gamma^\mu  \tau^I Q_d)
\end{equation}
which can help alleviate the possible tension with BR$(B_c\to \tau \nu)$ at the cost of a contribution to the decay $B\to K \nu \nu$. Such a model involving two leptoquarks can be motivated by a neutrino mass model. In this section we sketch out how this is possible leaving a detailed analysis to future work.

%%%%%%%%%%%%%%%%%%%%%%%%%%%%%%%%%%%%%%%%%%%%%%%%
\subsection{Neutrino masses}\label{Sec:neutrino}
\begin{figure}[tbh]
\centering
\includegraphics{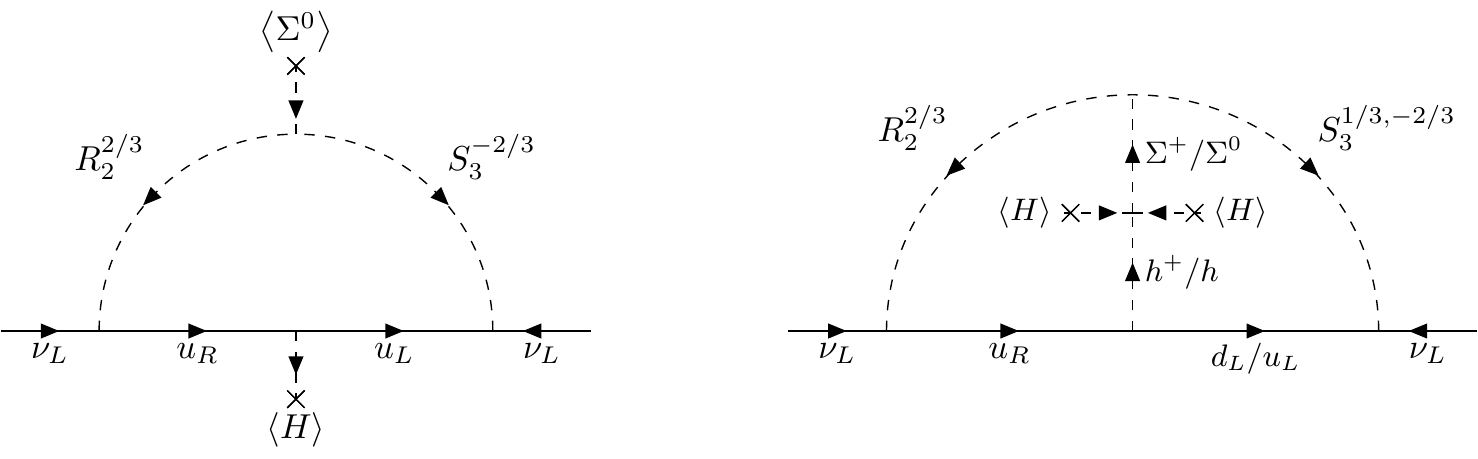}
\caption{Neutrino mass from leptoquarks in the loop. The superscripts $q$ of $R_2^q$ and $S_3^q$ denote the electromagnetic charge of the different components of $R_2$ and $S_3$, respectively.}
\label{fig:mnu}
\end{figure}

Just extending the SM with $R_2$ and $S_3$ leptoquarks is not sufficient to generate nonzero neutrino masses.\footnote{See Ref.~\cite{Dorsner:2017wwn} for a discussion of different possibilities in the context of a grand unified theory.} To keep our model minimal we extend our two leptoquark extension of the SM by a single particle which is a SU(2)$_L$ quadruplet with quantum numbers $\Sigma\sim(1,4,\frac{3}{2})$. Then neutrino masses are generated at the 1-loop level as shown in the left panel of Fig.~\ref{fig:mnu}. There is also a 2-loop contribution that is shown on the right.

If the 2-loop diagram can be neglected,
 neutrino masses are approximately given by their 1-loop contribution\footnote{Further details we relegate to the appendix.}
\begin{equation}
	(m_{\nu})_{ij}\simeq\frac{1}{16\pi^2}\frac{\mu \langle\Sigma^0\rangle}{m_{R_2}^2-m_{S}^2} \left(Y_{2}\right)_{ki}\left\{m_{k} \left[F\left(\frac{m_{R_2}^2}{m_{k}^2}\right)-F\left(\frac{m_{S}^2}{m_{k}^2}\right)\right] V_{kl}^*\right\} \left(Y_{S}\right)_{lj} + (i\leftrightarrow j)
\end{equation}
in the limit of small mixing between $R_2$ and $S_3$ leptoquarks, which is generated by the trilinear potential term $\mu \Sigma^{* ijk}R_{2i}S_{3jk}$. The Yukawa coupling matrices are defined in the basis where charged lepton and down-type quark mass matrices are diagonal. Thus the loop diagram with up-type quarks in the loop is proportional to the CKM mixing matrix element $V_{kl}^*$. Roman indices $i,j,k,l$ indicate flavor, $m_k$ the up-type quark mass, $m_{R_2},m_S$ are $R_2$ and $S_3$ leptoquark masses respectively, and $\langle\Sigma^0\rangle$ the vacuum expectation value of the neutral component of $\Sigma$. The loop function $F(x)$ is defined as
\begin{align}
	F(x)&=\frac{x\,\ln x}{1-x}
\end{align}
The more general expression for a general mixing angle between the $R_2$ and $S_3$ leptoquarks is given in appendix~\ref{sec:mnu_gen}. The 2-loop contribution features a similar flavor structure.

\section{Conclusion}\label{Sec:conclusion}
We demonstrate that the $R_2\sim(3,2,7/6)$ leptoquark is a new single particle candidate for explaining the anomalous lepton-flavor-universality ratios $R_{K^{(*)}}$ and $R_{D^{(*)}}$.
There is possibly a mild tension with the theoretically derived limit on the branching ratio BR($B_c \to \tau \nu$). Since we require the branching ratio of $B_c \to \tau \nu$ to be within a relatively narrow range, the viability of the $R_2$ leptoquark as a single particle solution to these anomalies is a directly falsifiable scenario. Another promising probe of the viable parameter space of our model is BR$(\tau \to e \gamma )$ where the projected sensitivity for Belle II is expected to improve by an order of magnitude~\cite{Kou:2018nap}. Another intriguing possibility is whether the large imaginary couplings needed to explain $R_{D^{(*)}}$ leave an permanent neutron electric dipole moment that is detectable in future experiments \cite{Dekens:2018bci}.

The tension with the disputed aggressive limit on BR($B_c \to \tau \nu$) can be alleviated through the introduction of a $S_3$ leptoquark which can be motivated by a neutrino mass model as discussed in Sec.~\ref{sec:mnu_gen}. This suggests that even if future analysis indeed rules out the $R_2$ leptoquark as a single leptoquark solution to anomalous $B$ decays, it still can play a substantial role in an extended model. 

\section*{Acknowledgments}

We thank John Gargalionis and David Straub for useful discussions. We thank Andrew Akeroyd and Nejc Kosnik for useful comments to the first version and Monika Blanke for pointing out a typo in the first version. OP is supported by the National Research Foundation of Korea Grants No. 2017K1A3A7A09016430 and No. 2017R1A2B4006338.
TRIUMF receives federal funding via a contribution agreement with the National Research Council of Canada and the Natural Science and Engineering Research Council of Canada.
This research includes computations using the computational cluster Katana supported by Research Technology Services at UNSW Sydney.

\appendix

\section{Leptoquark mixing and neutrino masses}
\label{sec:mnu_gen}
The relevant terms in the scalar potential $V=V_0+V_1$ are
\begin{align}
    V_{0} &=\displaystyle\sum_{\substack{H,R_2,S_3,\\ \Sigma\in \text{x}}} \left(\left(-1\right)^{q_\text{x}}\mu^2_{\text{x}}\left|\text{x}\right|^2+\frac{\lambda_\text{x}}{2}\left|\text{x}\right|^4\right)+\displaystyle\sum_{\substack{H,R_2,S_3,\\ \Sigma\in\left\{\text{x}< \text{y}\right\}}}\lambda_{\text{x}\text{y}}\left|\text{x}\right|^2 \left|\text{y}\right|^2\\
V_1&=\mu \Sigma^{* ijk}R_{2i}S_{3jk}+\lambda_{3\Sigma H}\Sigma^{* ijk}H_i H_j H_k+ \lambda_{2H33}R_2^{* i}S_{3 ij} S_{3 kl} H_m \epsilon^{jk} \epsilon^{lm}+ \text{h.c.}
\end{align}
where $(-1)^{q_x}$ is $-1$ for $H$ and $\Sigma$ and $+1$ for the other scalar fields.
Thus the general form for neutrino masses at 1-loop order is given by
\begin{equation}
	(m_{\nu})_{mn}=\frac{1}{16\pi^2}\left(U_s^{\dagger}\right)_{R_2 s_i} \left(Y_{2}\right)_{km}\left[m_k F\left(\frac{m_{S_i}^2}{m_k^2}\right) V_{kl}^*\right]^{kl} \left(Y_{S}\right)_{ln}\left(U_s\right)_{s_i S_3} + (m\leftrightarrow n).
\end{equation}
The mixing between $R_2^{2/3}$ and $S_3^{*2/3}$ is generated by $\left\langle\Sigma^0\right\rangle$ and is obtained by diagonalizing the charge $2/3$ leptoquark mass matrix which is given in the $(R_2^{2/3},S_3^{*2/3})$ basis
\begin{align}
    M_S^2&=\left(\begin{matrix}
    \mu_{R}^2 + \lambda_{HR}\frac{v_H^2}{2} + \lambda_{R\Sigma}\frac{v_{\Sigma}^2}{2} & \mu \frac{v_{\Sigma}}{\sqrt{2}} \\
    \mu \frac{v_{\Sigma}}{\sqrt{2}} & \mu_{S}^2 + \lambda_{HS}\frac{v_H^2}{2} + \lambda_{S\Sigma}\frac{v_{\Sigma}^2}{2}
    \end{matrix}\right)
    =U_S^T\text{Diag}\left(m_{S_1}^2,m_{S_2}^2\right) U_S 
\end{align}
with $v_H=\langle H^0\rangle/\sqrt{2}$ and $v_\Sigma=\langle \Sigma^0\rangle/\sqrt{2}$, the masses $m_{S_i}$ and the $2\times 2$ unitary mixing matrix $U_S$ which defines the mass eigenstates $S_i$ in terms of the flavor eigenstates
\begin{equation}
	\left(\begin{matrix}S_{1} \\ S_{2}\end{matrix}\right)
	=U_S
	\left(\begin{matrix}R_2^{2/3} \\ S_3^{*2/3}\end{matrix}\right)
	\qquad\qquad
    U_S=\begin{pmatrix}
		\cos\theta & \sin\theta\\
		-\sin\theta & \cos\theta
	\end{pmatrix}\;.
\end{equation}
A straightforward calculation results in the following expressions for the rotation angle and the masses
\begin{align}
    \tan\left(2\theta\right)&=\frac{\sqrt{2} \mu v_{\Sigma}}{\mu_{R}^2-\mu_{S}^2+\frac{v_H^2}{2}\left(\lambda_{HR}-\lambda_{HS}\right)+\frac{v_{\Sigma}^2}{2}\left(\lambda_{R\Sigma}-\lambda_{S\Sigma}\right)}\\
    m_{S_{1,2}}^2&=\frac{\mu_{R}^2+\mu_{S}^2+\frac{v_H^2}{2}\left(\lambda_{HR}+\lambda_{HS}\right)+\frac{v_{\Sigma}^2}{2}\left(\lambda_{R\Sigma}+\lambda_{S\Sigma}\right)}{2}\\\nonumber
    &\pm\frac{1}{2}\sqrt{\left[\mu_{R}^2-\mu_{S}^2+\frac{v_H^2}{2}\left(\lambda_{HR}-\lambda_{HS}\right)+\frac{v_{\Sigma}^2}{2}\left(\lambda_{R\Sigma}-\lambda_{S\Sigma}\right)\right]^2+4\mu^2\frac{v_{\Sigma}^2}{2}}\;.
\end{align}
For small $\mu$ and thus small mixing, the square of the masses $m_{R_2}$ and $m_S$ in the main part of the text can be identified with the diagonal elements of the scalar mass matrix $M_S^2$, $m_{R_2}^2=\mu_R^2+\lambda_{HR} v_H^2/2 + \lambda_{R\Sigma} v_\Sigma^2/2$ and $m_S^2=\mu_S^2+\lambda_{HS}v_H^2/2+\lambda_{S\Sigma}v_\Sigma^2/2$.

\bibliography{references}
\end{document}